\documentstyle[epsf]{l-aa}

\newcommand{\D}{\Delta}

\newcommand{\e}{\eta}
\newcommand{\k}{\kappa}
\newcommand{\ti}{\tilde}
\newcommand{\eqb}{\begin{eqnarray}}
\newcommand{\eqe}{\end{eqnarray}}
\newcommand{\diff}{{\rm d}}

\begin{document}
 
\thesaurus{12(02.01.1;02.19.1;02.18.5,10.03.1;11.02.1)}

\title{COMPUTATION OF DIFFUSIVE SHOCK ACCELERATION USING STOCHASTIC DIFFERENTIAL 
EQUATIONS}

\author{A. Marcowith 
\and J. G. Kirk}
 
\offprints{A. Marcowith at e-mail address marcowit@levi.mpi-hd.mpg.de}
 
\institute{Max-Planck-Institut f{\"u}r Kernphysik, Postfach 10~39~80, D-69029,
Heidelberg, Germany}

\date{Received..........; accepted..........}
\maketitle
 
\markboth{Marcowith \& Kirk: Computation of diffusive shock ...}{}

\begin{abstract}
The present work considers diffusive shock acceleration at 
non-relativistic shocks using a system of stochastic differential 
equations (SDE) equivalent to the Fokker-Planck equation. We 
compute approximate solutions of the transport of cosmic particles
at shock fronts with a SDE numerical scheme. Only the first order Fermi
process is considered. The momentum gain is given by implicit 
calculations of the fluid velocity gradients using a linear interpolation 
between two consecutive time steps. We first validate our procedure in 
the case of single shock acceleration and retrieve previous analytical 
derivations of the spectral index for different values of the P\'eclet 
number. The spectral steepening effect by synchrotron losses is also 
reproduced. A comparative discussion of implicit and explicit schemes 
for different shock thickness shows that implicit calculations extend 
the range of applicability of SDE schemes to infinitely thin 1D shocks. 
The method is then applied to multiple shock acceleration that can be 
relevant for Blazar jets and accretion disks and for galactic centre sources. 
We only consider a system of identical shocks which free parameters 
are the distance between two consecutive shocks, the synchrotron losses 
time and the escape time of the particles. The stationary distribution 
reproduces quite well the flat differential logarithm energy distribution 
produced by multiple shock effect, and also the piling-up effect due 
synchrotron losses at a momentum where they equilibrate the acceleration rate. 
At higher momenta particle losses dominate and the spectrum drops. 
The competition between acceleration and loss effects leads to a pile-up shaped 
distribution which appears to be effective only in a restrict 
range of inter-shock distances of $\sim$ 10-100 diffusion lengths. We finally 
compute the optically thin synchrotron spectrum produced such periodic 
pattern which can explain flat and/or inverted spectra observed in 
Flat Radio spectrum Quasars and in the galactic centre.

\end{abstract}

\section{Introduction}
\label{introduction}
The non-thermal radio and high-energy radiation spectra from 
extragalactic and galactic objects such as X-ray binaries, micro-Quasars, 
active galactic nuclei, jets and gamma-ray bursts require the presence of 
non-thermal particle distributions. One of the prime mechanisms for producing 
such distributions is diffusive acceleration at a shock front.
This theory assumes the particle distributions are isotropised by efficient 
scattering on wave turbulence on both sides of the shock and gain energy by a
first order Fermi process upon crossing it. The equation governing the particle
distribution is of the Fokker-Planck type, containing both advection (dynamical
friction) and diffusion terms in the spatial variables and an advection
term in the energy variable (or in the magnitude of the particle momentum $p$), 
which is proportional to the divergence of the fluid flow, and remains 
valid in an integral sense even across shock fronts. Many generalisations of this 
equation have been discussed, but for the application to non-thermal radiation 
spectra, the most relevant are the inclusion of synchrotron losses 
(Webb et al~\cite{webbetal84}) and the extension to systems containing multiple 
shock fronts (Blandford \& Ostriker~\cite{blandfordostriker80}, 
Spruit~\cite{spruit88}, Achterberg~\cite{achterberg90}, Schneider~\cite
{schneider93}, Melrose~\cite{melrose96}, Melrose \& Crouch~\cite{melrosecrouch97}).
However, because of the difficulty of solving the Fokker-Planck equation when the
coefficients are complicated (and possibly discontinuous) functions of
position, these papers either adopted an idealised situation, or developed
approximation schemes valid in only part of the parameter space.
A particularly interesting alternative approach is to use the equivalence of 
the Fokker-Planck equation to a system of stochastic differential equations 
(SDE's). Numerical integration of these is then akin to a Monte-Carlo simulation 
of the problem, which is relatively simple to implement, applies to complex 
flow patterns and places no restriction on the number of phase-space dimensions 
(e.g., Gardiner~\cite{gardiner83}). Several conventional Monte-Carlo simulations 
of particle acceleration exist (for a review see Jones \& Ellison~\cite
{jonesellison91}), usually assuming a prescribed form of the mean free path
of the particle as a function of the particle rigidity and plasma density. 
They have the advantage of being able to describe the evolution of both 
the thermal
and suprathermal populations, as well as the non-linear back reaction of the 
non-thermal component on the shock profile. Relativistic shocks and large 
angle scattering
are also easy to include in such simulations (see Ellison et al.~\cite{ellison90}).
However, this approach always makes additional assumptions concerning the 
particle trajectory (e.g., that it is unperturbed between isotropising 
\lq collisions\rq) which go beyond the diffusion approximation. 
In comparison, the SDE approach 
adopts the diffusion approximation for test particle and is 
restricted to the transport of suprathermal particles.

It is only recently that SDE systems have been applied to astrophysical 
problems: in solar physics with the investigation of acceleration of fast 
electrons in the solar corona (MacKinnon \& Craig~\cite{mackinnoncraig91},
Conway et al.~\cite{conwayetal98}) and in space physics with ion acceleration 
at the solar termination shock (Chalov et al.~\cite{chalovetal95}). 
In two papers Achterberg \& Kr\"ulls~(\cite{achterbergkruells92}) and Kr\"ulls \&
Achterberg~(\cite{KA94} -- henceforth KA94), have 
applied the SDE approach to the problem of particle acceleration at 
astrophysical shocks, including the possibility of second-order Fermi
acceleration (Schlickeiser~\cite{schlickeiser89}). 
These authors drew several important conclusions. In particular, they showed 
that for Kolmogorov turbulence the second-order acceleration effect 
is restricted to a small momentum range close to that of injection, because 
of the increase of the diffusion time with momentum. At high momentum, 
in contrast, the spectrum is formed by the competing effects of synchrotron 
losses and the first-order Fermi process. 

The numerical scheme used in KA94 is {\it explicit}, that is, the first-order
Fermi acceleration term in the SDE is integrated forwards in time using the 
value of the fluid velocity gradient at the beginning of each step. 
This method requires a step short enough to resolve sharp features in the flow, 
such as a shock transition, and  thus severely limits the ability to simulate 
acceleration in complex flow patterns containing structure on a large range of 
spatial scales. Our purpose in this paper is to propose and test an 
{\it implicit} method of integrating the SDE's. By implicit we mean that the 
coefficient of the first-order Fermi acceleration term is computed by linear 
interpolation between the end points of each time step. 
Such a scheme allows one to treat discontinuous shock structures using a finite time
step and thus opens up the prospect of finding approximate numerical 
solutions of the advection-diffusion equation in complex flow patterns where
the gradients of the fluid flow may become large over a distance much shorter
than the shock thickness. This is our longer term goal; for the present, 
we limit ourselves to the problem of a 1D single shock or to systems of 
multiple shocks.
  
The organisation of the paper is as follows: in Sect.~\ref{sdesystem}
we present the SDE system equivalent to the advection-diffusion equation and 
describe the explicit and implicit integration methods. 
We then test the implicit scheme in Sect.~\ref{singleshock} by applying it 
to the (well-known) problem of acceleration at a single isolated shock front, 
both with and without synchrotron losses. We also compare the 
performances of implicit and explicit schemes. 
Section~\ref{multiple} focuses on the application to acceleration in a system of 
multiple shocks. Here, as well as testing the code against known approximate
solutions, we present new solutions valid in regions of parameter space 
inaccessible to the analytic methods. We use the results to calculate the
optically thin synchrotron radiation produced by the stationary particle 
distribution within a periodic pattern of shocks. In Sect.~\ref{conclusions}
we summarise our results and discuss extensions of the method to more
complex problems.

\section{Monte-Carlo simulations}
\label{sdesystem}
\subsection{Formulation of the SDE's}
The usual form of the advection-diffusion equation, describing the 
transport of cosmic-rays (Skilling~\cite{skilling75}) is (in 3D)
\eqb
\frac{\partial f}{\partial t} + \left(\vec{u}\cdot\nabla\right)f
-\frac{1}{3}\left(\nabla\cdot\vec{u} \right)p
\frac{\partial f}{\partial p} &=& S + D \,.
\label{skilling}
\eqe
where $S$ is a source term and $D$ is a diffusion operator of the form
\eqb
D &=& \nabla\left( \overline{\overline{\k}}\nabla f\right) \, .
\label{spatialdiff}
\eqe
The diffusion tensor $\overline{\overline{\k}}$ describes the spatial
transport of particles. Adding the effects of synchrotron losses, and
second-order Fermi acceleration  
one finds
\eqb
D&=&    
\nabla\left( \overline{\overline{\k}}\nabla f\right)
 +\frac{1}{p^2}\frac{\partial}{\partial p}\left[
a_2(\vec{x},p) p^2 \frac{\partial f}{\partial p}-\right.
\nonumber\\
&&\left.{\phantom f\over\phantom f}\left(\vec{a}_1(\vec{x},p)
\cdot\nabla\right) p^2 f  +a_{\rm s}(\vec{x})p^4 f\right] \, .    
\label{extended}
\eqe
Here the coefficients $a_1$ and $a_2$ describe
the second-order Fermi process and $a_{\rm s}$ synchrotron 
losses. Equations (\ref{skilling}) and (\ref{extended}) 
give the full advection-diffusion equation
of cosmic particles in the diffusion approximation.

The general system of SDEs describing the motion of a point $\vec{X}$
phase space can be written as:
\begin{equation}
\frac{d\vec{X}}{dt} = \vec{A}(\vec{X}) + \overline{\overline{B}}(\vec{X}) 
\frac{d\vec{W}_{\vec{X}}}{dt} \ ,
\label{sde1}
\end{equation}
where $W_{\vec{X}}$ is a Wiener 
process: a stochastic diffusive process used to describe Brownian motion, 
with a conditional probability which follows a Gaussian distribution.
For initial conditions given by $W_{\vec{X}}=W_{\vec{X}_0}$ at $t=t_{0}$, we
have at time $t$:
\begin{eqnarray}
\left<W_{\vec{X}}\right> &=& W_{\vec{X}_0} \ , 
\nonumber \\
\left<(W_{\vec{X}}-W_{\vec{X}_0})^2\right> &=& t-t_0 \ .
\end{eqnarray}
where $\left<\phantom)\right>$ means the average value.
It\^o~(\cite{ito51}) has shown that the distribution function describing the 
stochastic trajectories of the point $\vec{X}$ 
obeys the Fokker-Planck equation if the
coefficients $\vec{A}$ and $\overline{\overline{B}}$ are identified
with the dynamic friction and diffusion coefficients of this equation.
In the case of the Fokker-Planck equation (\ref{skilling}), together with 
(\ref{spatialdiff}) or (\ref{extended}), the phase space is 
four-dimensional, $\vec{X}=(\vec{x},p)$,
and expressions for the coefficients $\vec{A}$ and
$\overline{\overline{B}}$ are given by KA94.
In this paper we shall restrict ourselves to the case of one space 
dimension, and include only the terms in Eq.~(\ref{extended}) 
describing spatial diffusion and synchrotron losses, i.e., 
$a_{1,2}=0$, so that the phase space is two-dimensional: $\vec{X}=(x,p)$. 
Further simplifying to the case 
where the spatial diffusion coefficient is independent of both position and 
momentum and the loss rate $a_{\rm s}$ is independent of position, 
the set of SDE's (\ref{sde1}) reduces to
\eqb
\diff x&=& u\diff t + \sqrt{2\k} \  \diff W_{x} \ ,
\\
\diff \ln(p) &=&-\left( a_{\rm s} p +\frac{1}{3}\frac{\diff u}{\diff x}\right)
\diff t
\label{diffsde}
\eqe
In this system, $W_x$ is a continuous but non-differentiable 
process, so that, as written, these equations do not exist in a strictly
mathematical sense. 
To overcome this problem It\^o~(\cite{ito51}) (see also Gardiner~\cite{gardiner83})
has defined 
stochastic differential integrals of the form $\int_t B(X(t),t) dW_{X}$.
Approximate solutions of the system of SDE's can be found by discretizing 
in time:

\eqb
\Delta x\,=\,x_{k+1}-x_{k} &=& u(x_k) (t_{k+1}-t_k) + \sqrt{2\k}\times
\nonumber \\
&& \left[W(t_{k+1})-W(t_k)\right] \, 
\label{sde2}
\\
\ln\left(p_{k+1}/p_k\right)&=&
-\left(a_{\rm s}p_k +{1\over3}\left.{\diff u\over\diff x}\right|_k\right)
\left(t_{k+1}-t_k\right)
\label{sde3}
\eqe
The term in brackets is the increment of the Wiener process, which
is proportional to the square root of $t_{k+1}-t_k=\D t$: 
\eqb
\D W &=& \xi \D t^{1/2} \, ,
\eqe
where $\xi$ is a Gaussian distributed random number with zero mean and 
unit variance.
This is the so-called {\it Cauchy-Euler} procedure (Gardiner~\cite{gardiner83}), 
as used by KA94. At each time step $\Delta t$, the change in $x$ is made up of 
two parts: an advective (and deterministic) step
\eqb
\Delta x_{\rm adv}&=& u(x_k)\Delta t
\eqe
and a diffusive (stochastic) step
\eqb
\Delta x_{\rm diff}&=& \xi\sqrt{2\k\Delta t} \,.
\eqe
The approximate solutions converge to the exact solution of the SDE for $\D t 
\rightarrow 0$. However, this scheme has the disadvantage that $\Delta t$ must be
small enough to resolve the spatial structure in $u(x)$.
Denoting by $x_{\rm shock}$ the shortest lengthscale associated with $u(x)$, KA94
found for a particular example the requirement
\eqb
\Delta x_{\rm adv}&\ll& {\rm Min}\left(\Delta x_{\rm diff}, x_{\rm shock}\right)
\label{eulercond}
\eqe
Thus, although the diffusive step can be long compared to $x_{\rm shock}$, the
advective step must resolve it, and the method is not appropriate for 
flow patterns containing sharp gradients (or discontinuities). 

\subsection{Implicit Euler schemes}
Implicit methods, in which the increments are expressed not
just in terms of the solution at the start of a time step, but implicitly in
terms of the solution at the end of it, are frequently effective in relieving 
time-step problems, and have been discussed for SDE's by 
Smith \& Gardiner~(\cite{smithgardiner89}). Their main advantage is that they 
yield stable algorithms. However, the problem raised by the condition
(\ref{eulercond}) is not one of instability, but accuracy. Furthermore,
it would appear that the advective, deterministic term is more sensitive to the
problem than is the diffusive term. 
In view of this, we have chosen to test an algorithm in which, for the advective 
term, the coefficient is evaluated neither at the initial point of a
time step (explicit) nor at the end point (fully implicit) but is integrated 
exactly over the step, using a linear interpolation of the trajectory. 
In this way, we may hope to account approximately for changes in the velocity 
$u(x)$ which occur on very small length scales, unresolved by either 
$\Delta x_{\rm adv}$, or $\Delta x_{\rm diff}$. 

Replacing the advective terms in Eqs.~(\ref{sde2}) and (\ref{sde3})
using 
\eqb
x&=&{ \Delta x\over \Delta t}t
\label{interpol}
\eqe
we find, neglecting for the moment the synchrotron losses,
\eqb
\Delta x&=& \int_{t_k}^{t_{k+1}}\diff t u(x) + \sqrt{2\k}
\left[W(t_{k+1})-W(t_k)\right] \, 
\nonumber\\
&=&{\Delta t\over\Delta x}\int_{x_k}^{x_{k+1}}\diff x \,u(x) 
\nonumber\\
& &+ \sqrt{2\k} \left[W(t_{k+1})-W(t_k)\right] \,
\label{impl1}
\\
\ln\left({p_{k+1}
\over
p_k}\right)&=&
-{1\over3}\int_{t_k}^{t_{k+1}}\diff t {\diff u\over\diff x}
\nonumber \\
&=&- {\Delta t\over3\Delta x}\left[u\left(x_{k+1}\right)-u\left(x_k\right)\right]\,,
\label{impl2}
\eqe
which has the character of an implicit scheme, since the right-hand side of 
Eq.~(\ref{impl1}) is a function of $x_{k+1}$. This
technique has been used in applications of SDE to other fields
(see Kl\"oden \& Platen~\cite{kloedenplaten92}), but we are not aware of a
detailed discussion of its properties. We show in the following that for the
test cases we have examined, the scheme yields accurate results when 
the condition (\ref{eulercond}) is replaced by the less restrictive
one:
\eqb
\Delta x_{\rm adv}&\ll& \Delta x_{\rm diff}
\label{diffcond}
\eqe


\section{The test case of acceleration at a single shock}
\label{singleshock}
We consider an infinite 1D plasma in which particles
propagate with diffusion coefficient $\kappa_{\rm D}$. The flow velocity of the
plasma is constant for $x<-x_{\rm shock}/2$ (the upstream region) and equal to
$u_1$ in the shock restframe. Similarly, in the downstream region, 
$x>x_{\rm shock}/2$, the velocity  is constant, $u_2=u_1/r$, where $r$ is 
the compression ratio of the shock. Position, time and momentum are normalised 
to the diffusion length and time scales upstream: ($\kappa_{\rm D}/u_1$, 
and $\kappa_{\rm D}/u_1^2$ respectively) and the injection momentum of a 
particle $p_{\rm i}$. Thus we define the following dimensionless variables and 
coefficients:
\eqb
\tilde{x}&=&x u_1/\kappa_{\rm D}
\nonumber\\
\tilde{p}&=&p/p_{\rm i}
\nonumber\\
\tilde{t}&=&tu_1^2/\kappa_{\rm D}
\nonumber\\
\tilde{u}&=&u/u_1
\nonumber\\
\tilde{a}_{\rm s}&=&a_{\rm s}(x)/a_{\rm s,0}
\nonumber\\
\eta&=& u_1\int_{-x_{\rm shock}/2}^{x_{\rm shock}/2}\diff x /\kappa_{\rm D}
\nonumber\\
&=&u_1 x_{\rm shock}/\kappa_{\rm D}
\end{eqnarray}
where $a_{\rm s,0}= 3 u_1^2/ (2 \kappa_{\rm D} p_i)$ 
controls the synchrotron losses, and $\eta$ is the P\'eclet
number, since the shock transition is confined to the region $|\tilde{x}|<\eta/2$.

Using the scheme (\ref{impl1},\ref{impl2}) we have simulated trajectories in 
shocks of different $\eta$. 
Each simulation runs with a given number of particles $N$, injected
at momentum $\tilde{p} = 1$, for a given computation time 
$\tilde{t}_{\rm max}$. Particles can escape at $\tilde{t}<\tilde{t}_{\rm max}$ 
by crossing a boundary at $|\tilde{x}|=L\gg1$. The particle distribution
is measured at the shock front. Each crossing with an initial momentum
$\tilde{p} \ge 1$ increases by unity the differential logarithmic
distribution in the bin of momentum bracketing $\tilde{p}$.

It is easily seen from Eq.~(\ref{impl2}) that for $\eta\ll1$ large values of
$\Delta\ln\tilde{p}$ occur if the advective and diffusive steps almost cancel:
$\Delta\tilde{x}_{\rm adv}\approx-\Delta\tilde{x}_{\rm diff}$. 
In this case the result is particularly sensitive to our assumption that the 
trajectory between the points $\tilde{x}_k$ and $\tilde{x}_{k+1}$ can 
be interpolated linearly. 
A partial solution 
to this problem is to reduce the time step which automatically 
decreases the probability of choosing a diffusive step which cancels the 
advective step. However, this procedure increases the computation time. 
The problem can be avoided completely by replacing the Wiener process by a 
different random process possessing the same mean and variance, but which 
effectively prevents the rare steps with $\Delta \tilde{x}_{\rm adv}\approx-
\Delta \tilde{x}_{\rm diff}$. For a large number of trials, the random step 
produced by any such distribution tends to the Gaussian form of $W$. The 
simplest choice is to take $\xi=\pm1$ with equal probability for each sign.

We first test this prescription in the most difficult case of infinitely thin 
shocks where $\eta=0$, in which case the profile shows a discontinuous change in 
velocity between the up and downstream regions. 

\begin{figure}[htbp]  
\centerline{\epsfxsize=9cm \epsfbox{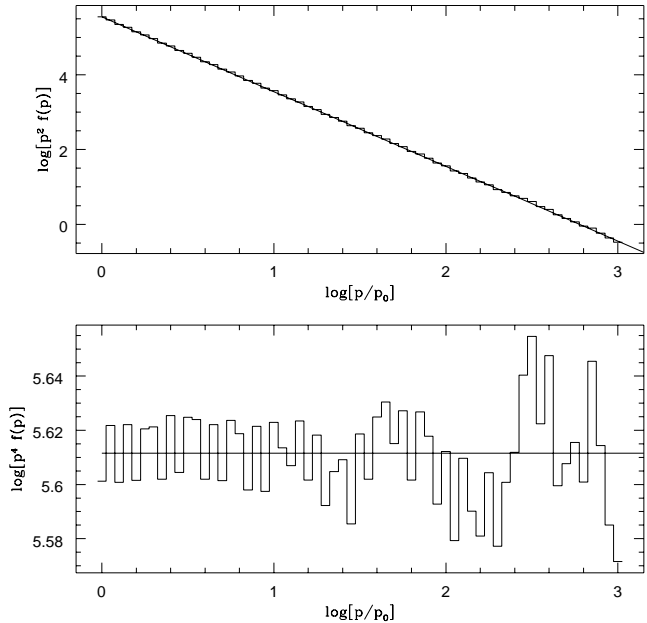}} 
\caption{
\label{test1}
Single shock stationary solution. The parameters are:
compression ratio $r=4$,
maximum trajectory time $\tilde{t}_{max} = 3000$, boundaries
$L = \pm 1000$, time step $\D \tilde{t} = 10^{-3}$. The $x$ step is calculated
implicitly using (\protect\ref{impl1}). {\it Upper panel}: comparison 
between numerical distribution and the analytic solution $f\propto 
\tilde{p}^{-4}$ (solid line). {\it Lower panel}: the distribution weighted with 
$\tilde{p}^4$.}
\end{figure}

Figure \ref{test1} gives the stationary spectrum obtained with the modified scheme 
described above. We found over at least two decades of energy a stationary 
power-law distribution function of index $s = 3r/(r-1) =4$ in the case of 
strong shocks. Larger computation times are needed to reach the same accuracy 
at larger momentum. A $\chi^{2}$ test against the analytic solution gives a value 
$\sim 1$.

\begin{figure}[htbp]  
\centerline{\epsfxsize=9cm \epsfbox{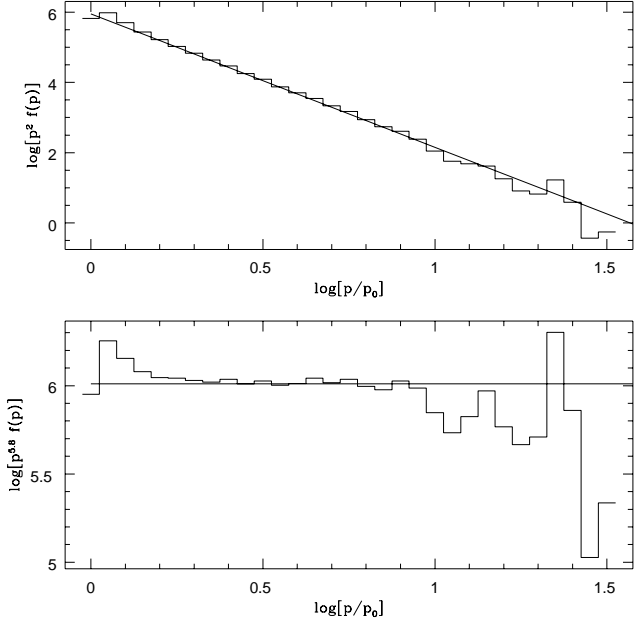}} 
\caption{
\label{test2}
{\it Upper panel}: the distribution for same parameters as in 
Fig.~(\protect\ref{test1}),
but with the P\'eclet number $\eta = 10$ and time step $\D \tilde{t} = 10^{-3}$, 
compared to the analytical solution $f\propto \tilde{p}^{-5.8}$.
{\it Lower panel}: the spectrum for single shock weighted by $\tilde{p}^{5.8}$.}
\end{figure}

We consider now the case of a shock with P\'eclet
number $\e \ge 1$. We compare our results with the semi-analytical 
derivation of the spectral index by 
Schneider \& Kirk~(\cite{schneiderkirk87}).
The simulations here use a linear velocity profile 
\eqb
\tilde{u}(\tilde{x})&=&{1+r\over2r} -{(r-1)\tilde{x}\over
\eta r}
\eqe
For larger P\'eclet number, the particles experience
a smaller velocity jump at each step. The Fermi process is thus less efficient,
leading to softer stationary distributions.
For a P\'eclet number of $\e = 10$ we get an index of $\sim 5.8$ 
(see Fig. \ref{test2}) in good agreements with the results of 
Schneider \& Kirk (\cite{schneiderkirk87}). The discrepancies seen 
at small momenta are not statistical errors, but arise because our method 
assumes zero particle flux in space at the injection point, whereas the 
scale-independent power-law distribution implies a finite diffusive flux. 
This \lq transient\rq\ effect disappears at momenta slightly above that of 
injection. 

\subsection{Synchrotron losses}
Synchrotron losses in the diffusive shock acceleration process 
have been investigated analytically by 
Webb et al.~(\cite{webbetal84}). Their main effect 
is to soften the spectrum at momenta greater than $\tilde{p}_*$ 
where the loss  rate equals the acceleration rate.
The inclusion of loss terms in the scheme modifies the way the momentum gain 
is calculated. Returning to Eq. (\ref{diffsde}), and using the linear 
interpolation of the trajectory (\ref{interpol}) we arrive at the ordinary 
differential equation
\eqb
{d\tilde{p}\over d\tilde{x}}&=& - \tilde{a}_{\rm s} p^2 
{\Delta \tilde{t}\over\Delta \tilde{x}}-{1\over3}
{\diff \tilde{u}\over\diff\tilde{x}} {\Delta \tilde{t}\over\Delta \tilde{x}}
\tilde{p} \, .
\eqe
For the initial condition $\tilde{p}=\tilde{p}_k$ at $\tilde{x}=\tilde{x}_k$ 
(i.e., $\tilde{t}=\tilde{t}_k$) 
the solution is
\eqb
\ln\left(\tilde{p}/\tilde{p}_k\right) &=& -\ln\left( F_{\rm I}+L_{\rm s}\right) \, ,
\label{solution}
\eqe
where the first-order Fermi term ($F_{\rm I}$) and the loss 
term $L_{\rm s}$ are
\eqb
F_{\rm I}&=& \exp\left({\Delta \tilde{u}\over3 \Delta \tilde{x}}\Delta \tilde{t}
\right) \ ,
\eqe
and 
\eqb
L_{\rm s} &=& \tilde{a}_{\rm s} {\Delta \tilde{t}\over\Delta \tilde{x}} 
\tilde{p}_k \int_{\tilde{x}_k}^{\tilde{x}} 
\exp\left(-{\Delta \tilde{u}\over3 \Delta \tilde{x}}\Delta \tilde{t}\right) 
\diff \tilde{x}' \, .
\eqe
This solution is exact for the linearly interpolated trajectory (\ref{interpol}).
Having determined $\tilde{x}_{k+1}$ from Eq. (\ref{impl1}), the new value of 
the momentum is given by Eq. (\ref{solution}) with $\tilde{x}=\tilde{x}_{k+1}$.

\begin{figure}[htbp]  
\centerline{\epsfxsize=9cm \epsfbox{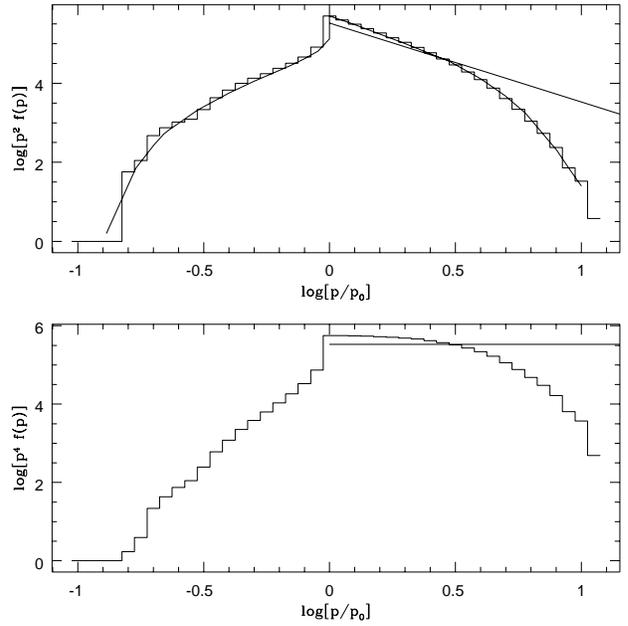}} 
\caption{
\label{lossfig}
{\it Upper panel}: the distribution 
for acceleration at a single shock front including synchrotron losses for 
$a_{{\rm s}1}=10$, $a_{{\rm s}2}=1$ and $r=4$.
The thin solid line is adapted from the analytical solution given by
Webb et al.~(\protect\cite{webbetal84}). 
The maximum time for a single trajectory is $\tilde{t}_{max} = 3000$, and 
$\Delta \tilde{t} = 10^{-3}$.
{\it Lower panel}: the distribution weighted by $\tilde{p}^4$. In each case,
the solid line corresponds to the loss free solution.}
\end{figure}

In Fig. \ref{lossfig}, we show the results of this implicit scheme.
Following Webb et al.~(\cite{webbetal84}) and KA94, 
we define upstream  and downstream synchrotron coefficients 
($a_{\rm s1,s2}$), given by $a_{{\rm s}i} = 
\tilde{u}^2/(4 \ti{\k}\ti{a}_{s} \tilde{p}_{\rm i})$. This gives for the 
characteristic momentum $\tilde{p}_* = 4/(q/a_{{\rm s}1} + 
(q-3)/a_{{\rm s}2})$, with $q = 3r/(r-1)$. For a compression ratio $r = 4$, we have 
$a_{s1} = 10$ and $a_{s2} = 1$, and the distribution cuts-off
for $\tilde{p}> \tilde{p}_* = 2.9 $ (i.e., $\log(\tilde{p})>0.46$). 
The solution showed in figure 3 is in good agreement with the analytical result 
of Webb et al.~(\cite{webbetal84}). 

The implicit scheme is, of course, much faster than the explicit scheme for 
flows with small P\'eclet numbers, since the latter are accurate only when
$\Delta\tilde{t}\ll \eta$, whereas for the former $\Delta\tilde{t}\ll1$ 
suffices (see Table~\ref{table}). 
However, it is also interesting to compare the implicit and explicit schemes 
in flows with large P\'eclet numbers, where both should be accurate. 
In this case, the explicit and implicit schemes give similar results 
for single shocks for $\Delta \tilde{t}\sim 10^{-3}-10^{-2}$. For larger time steps 
the particles tend to be advected prematurely from the acceleration zone and 
both schemes show an unphysical softening of the spectrum.
For a given $\Delta \tilde{t}$, the explicit scheme is faster by a factor 
$\sim 5/3$, since all quantities entering into the computation of the position 
and momentum increments are calculated only once per time step.

\begin{table}
\caption{
\protect\label{table}
Summary of the results of the explicit and implicit schemes. 
The power-law index of the distribution is compared with the value given by an
analytic calculation (Schneider \& Kirk~\protect\cite{schneiderkirk87}).The typical
error on the index estimates is of order of $\sim {\rm few} \%$.}
\begin{tabular}{|l|l|l|l|l|} 
\hline 
$\eta$        &Analytic &$\Delta\tilde{t}$    &implicit    & explicit   \\ \hline
$10^{-12}$    &$s=4.0$  &$10^{-3}$     & s=4.0      &  -----    \\ 
$2 10^{-3}$   &$s=4.0$  &$10^{-3}$     & s=4.01     & s=3.98    \\
$2 10^{-3}$   &$s=4.0$  &$10^{-2}$     & s=3.99     & s=4.55    \\
$2 10^{-3}$   &$s=4.0$  &$10^{-1}$     & s=4.04     &  -----    \\
$1$           &$s=4.18$ &$10^{-3}$     & s=4.18     & s=4.19    \\
$1$           &$s=4.18$ &$10^{-2}$     & s=4.18     & s=4.19    \\
$1$           &$s=4.18$ &$10^{-1}$     & s=4.21     & s=4.21    \\
$10$          &$s=5.80$ &$10^{-3}$     & s=5.80     & s=5.79    \\
$10$          &$s=5.80$ &$10^{-2}$     & s=5.81     & s=5.82    \\
$10$          &$s=5.80$ &$10^{-1}$     & s=5.85     & s=5.92    \\\hline
\end{tabular}
\end{table}

\section{Acceleration at multiple shocks}
\label{multiple}
The subject of multiple shock acceleration has been extensively investigated
both analytically and numerically over the past few years (Spruit~\cite{spruit88}, 
Achterberg~\cite{achterberg90}, Schneider~\cite{schneider93}, 
Pope \& Melrose~\cite{popemelrose93}, 
Melrose \& Crouch~\cite{melrosecrouch97}). Analytic solutions of the 
diffusion-advection equation including synchrotron losses can be derived
if we assume that the time spent by a fluid element
between two consecutive shocks is much longer than the 
acceleration time at a single shock problem (see 
Schneider~\cite{schneider93}). 
Another way of formulating this condition is that the first-order Fermi 
acceleration process should be much faster 
than all other processes, such as escape, decompression and losses. 
With this hypothesis, a final power-law 
index can be calculated which takes account of the 
different generations of particles accelerated at individual shocks. 
At sufficiently high momentum, this approximation fails, since the 
time-scale of the losses must eventually become comparable to the acceleration 
time-scale. The effect of multiple shocks is to increase the acceleration 
efficiency, by reducing the effective escape rate. In a purely 1D system, 
the escape rate is formally zero, and the distribution functions tends to 
$f\propto p^{-3}$. 
As shown, using spatially averaged equations, 
by Kardashev~(\cite{kardashev62}) and by  Schlickeiser~(\cite{schlickeiser84}), 
in the presence of losses, this spectrum extends up to 
momentum values where the loss effect becomes dominant, and the spectrum piles 
up at a momentum where the effective acceleration time equals the loss time.
More recently, Protheroe \& Stanev~(\cite{protheroestanev99}) 
(see, however, Drury et al~\cite{druryetal99}) have 
proposed an alternative method of computing the cut-off and pile-up effects in 
the high energy particle spectrum with various energy dependent 
diffusion coefficients. They present both an analytical model and a 
conventional Monte-Carlo
simulation (of the same kind as described in the introduction) and show that 
the effect of the Klein-Nishina regime of inverse Compton losses may modify 
the results obtained with continuous (synchrotron and inverse Compton 
in the Thomson regime) losses. In this paper we include neither 
non-continuous energy losses nor energy dependent diffusion, but postpone
an investigation of these effects to future work.

In general for multiple shock system, the picture is somewhat more complicated
than depicted in previous analytical or semi-analytical models, since
the shocks may be so close together that the acceleration time at a single 
shock is not short compared to the flow time between the shocks. Also, 
at high momenta, the synchrotron loss time-scale must be compared not only to 
the acceleration time at a single shock, but also to the flow time between the 
shocks. 

\begin{figure}[htbp]  
\centerline{\epsfxsize=8cm \epsfbox{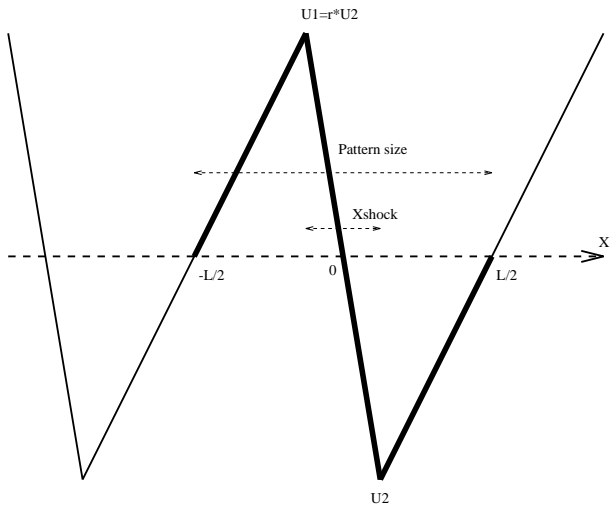}} 
\caption{
\label{pattern}
Geometry of the multiple shock system -- a single period of the 
pattern is drawn in bold.}
\end{figure}

To investigate these situations, we consider a simple periodic pattern 
(of period $L$) including shock fronts and re-expansion 
regions (see Fig. \ref{pattern}). The flow speed of the pattern 
($\tilde{x} \in [-L/2,L/2]$) can then be written as
\eqb
\tilde{u}(x)&=& \tilde{u}_1 + 
\left(\tilde{u}_1-\tilde{u}_2\right)\frac{\tilde{x}+x_{\rm s}/2}{L-x_{\rm s}} 
\, , \tilde{x} \in [-L/2,-x_{\rm s}/2] \, , 
\nonumber \\
\tilde{u}(x)&=& \tilde{u_1} - \left(\tilde{u}_1-\tilde{u}_2\right)
\frac{\tilde{x}+x_{\rm s}/2}{x_{\rm s}} 
\, , \tilde{x} \in [-x_{\rm s}/2,x_{\rm s}/2] \, , 
\nonumber \\
\tilde{u}(x)&=& \tilde{u}_2 + 
\left(\tilde{u}_1-\tilde{u}_2\right)\frac{\tilde{x}-x_{\rm s}/2}{L-x_{\rm s}} 
\, , \tilde{x} \in [x_{\rm s}/2,L/2] \, 
\label{patterndef}
\eqe
where $\tilde{u}_2=\tilde{u}_1/r$ is the flow speed on the downstream side of 
the shock.

There are two free parameters (for a given compression ratio) in this problem: 
the {\em advection time} $\tilde{t}_{\rm adv}$, 
which is the (dimensionless) time for the fluid to flow through one 
wavelength of the pattern and is numerically equal to the dimensionless 
length $L$, and the {\em loss strength} given by $\tilde{a}_{\rm s}$, 
which effectively defines a momentum scale, since the (dimensionless) synchrotron 
loss time at momentum $\tilde{p}$ is 
\eqb
\tilde{t}_{\rm syn}&=&{2\over 3\tilde{a}_{\rm s}\tilde{p}}
\label{synchlosstime}
\eqe
To these we add a third: an {\em escape time} $\tilde{t}_{\rm esc}$, assumed
independent of $x$ and $p$. This can be understood as a crude way of incorporating
2 or 3-dimensional effects into our 1-dimensional simulation, since 
only in 1 dimension are the particles unable to leave the train of shock fronts.
The inclusion of escape effects in the SDE system can be effected by 
rescaling the number of particles that have crossed the shock with at 
time $\tilde{t}$ by the factor $\exp(-\tilde{t}/\tilde{t}_{\rm esc})$.

There are then two important characteristic times which are important for 
a discussion of the solutions:
\begin{itemize}
\item
the single shock acceleration time $\tilde{t}_{\rm S}$, which is the characteristic 
time of momentum increase at an isolated shock front:
\eqb
\tilde{t}_{\rm S}&=&3r \frac{1+r}{r-1} \ ,
\label{lagage}
\eqe
(Lagage \& Cesarsky~\cite{lagagecesarsky81})
with $\tilde{\kappa} = \tilde{\kappa_1} = \tilde{\kappa_2} = 1$

\item the effective multiple shock acceleration time $\tilde{t}_{\rm M}$ 
which is controlled by both the advective time and the single shock acceleration 
time and is inferred from our numerical results.
\end{itemize}

We present, in the following, results for a shock 
system described by Eq.~(\ref{patterndef}) with a P\'eclet number $\eta = 0$. 
We adopt as typical values of the parameters:
compression ratio $r=4$, 
escape time $\tilde{t}_{\rm esc} = 10^3$, an inter-shock distance $L=20$, and, 
in the case of synchrotron losses, $\tilde{a}_{\rm s} = 6.5\times 10^{-6}$
(chosen to give a peak momentum of $10^3$).
All the simulations are run for a time of $\tilde{t}_{\rm max} = 5 
\tilde{t}_{\rm esc}$ and with a time step $\Delta\tilde{t} = 10^{-2}$. 

\subsection{Multiple shock effect: the case without losses}
We first consider the case where losses are inefficient ($\ti{a}_s \rightarrow
0$). The results are shown in Fig.~\ref{nmulti}a. The stationary solution at each 
shock front $f(x=nL,p)$ (with $n$ an integer) is close to $p^{-3}$. 
This confirms the multiple shock effect as an efficient way of producing 
higher energy particles and harder spectra than isolated shocks. 
In fact, the stationary index is not exactly $3$ but 
slightly steeper, owing to the finite escape time. Using the general relation
between the power-law index of accelerated particles and 
the escape and acceleration times: $f\propto p^{-s}$, with 
$s=3+\tilde{t}_{\rm acc}/\tilde{t}_{\rm esc}$ (Kirk et al.~\cite{kirketal94})
we find for the effective acceleration time in this particular multiple shock
system $\tilde{t}_{\rm M}\approx 10^2$. Thus, $\tilde{t}_{\rm M}$ is 
($\equiv L$) in this case.  

For momenta lower than that of injection, $\tilde{p} \le 1$, 
the method of Schneider (\cite{schneider93}), and
Melrose \& Crouch (\cite{melrosecrouch97}), gives a continuation of the 
power-law down to $p \sim 1/r$. This is clearly an artifact of the
assumed separation of the acceleration and expansion processes. It does not 
appear in our simulations, which show instead a hardening to lower frequencies 
starting at the point $\tilde{p}=1$.  However, because of the transients 
associated with our method close to the injection momentum, this effect may also 
be an artifact. 

At still lower momentum values, the analytical solution $f(p)\propto 
p^{\lambda}$ for an escape probability independent of momentum is given by 
(see equation 4.5 in Schneider~\cite{schneider93})
\begin{equation}
\frac{(3r/(r-1)-3)(r^{1+\lambda/3})}{3r/(r-1)+\lambda} = 1 + 
\frac{\tilde{t}_{\rm adv}}{\tilde{t}_{\rm esc}} \ .
\label{schneidereq}
\end{equation}
The stationary index is $\lambda = 0$ for $\tilde{t}_{\rm esc} \gg 
\tilde{t}_{\rm adv}$, but for lower values of the ratio of escape to multiple 
shock acceleration time, the spectrum hardens, and typically for a ratio 
$\tilde{t}_{\rm adv}/\tilde{t}_{\rm esc} \sim 0.1$, we get $\lambda 
\sim 0.2$, in good agreement (given the accuracy of the 
time-scales) with the simulations; for $p \le 1$ $f(p) \sim p^{0.3}$ down to 
$p =0.1$. 

\subsubsection{Variations of the escape time}
If the inter-shock distance $L$ is kept constant, increasing (decreasing) 
the escape time leads to a harder (softer) stationary spectra. 
This is clearly seen in Fig.~\ref{nmulti}b where we have reduced the escape 
time by a factor of 2 compared to the fiducial case of Fig.~\ref{nmulti}. 
The resulting spectrum has an index of $3.2$, which again gives
$\tilde{t}_{\rm M} = {\rm few} x \tilde{t}_{\rm adv}$. 
For momentum $\tilde{p} \le 1$, from Eq.~(\ref{schneidereq}), the relation 
$\tilde{t}_{\rm esc} \sim 5 \tilde{t}_{\rm M}$ gives a spectrum with 
$\lambda \sim 0.4$. We obtained down to $\tilde{p}=0.1$ an index of $\approx 0.5$.
If $\tilde{t}_{\rm esc} \le \tilde{t}_{\rm adv}$, the particles escape the system 
before being advected to the next shock. The stationary solution tends to 
the single shock result of Sect.~2.1 without losses, which is
a power-law spectrum with an index of $4$ (for $r=4$). 

\subsubsection{Variations of the inter-shock distance}
\begin{figure}[htbp]  
\centerline{\epsfxsize=10cm \epsfbox{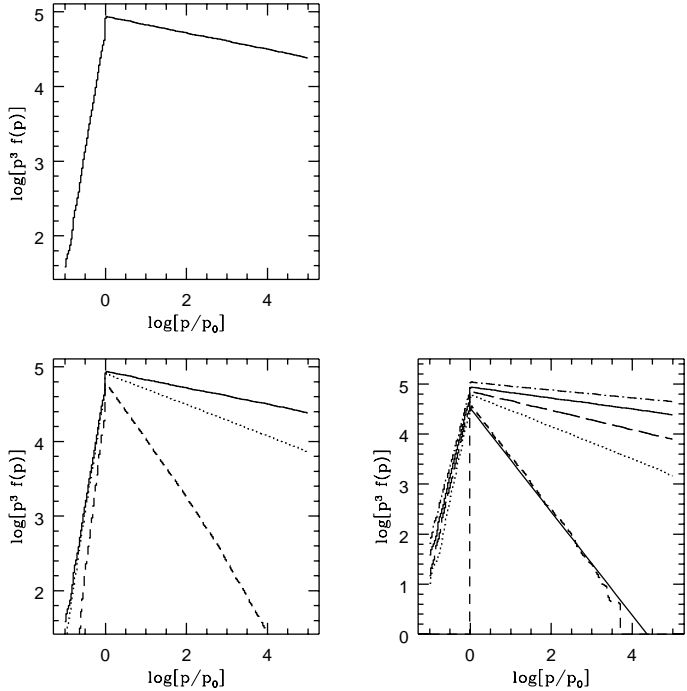}} 
\caption{
\label{nmulti}
Differential logarithm spectrum produced by an ensemble of identical 
shocks. {\it Upper left panel} (a): our refering case with 
$\tilde{t}_M/\tilde{t}_{esc} \sim 0.1$ (see text for parameters). 
{\it Lower left panel} (b): the case with different 
escape times, in solid line $\tilde{t}_{\rm esc}=1000$, 
in dotted line $\tilde{t}_{\rm esc} = 500$, and in short-dashed line 
$\tilde{t}_{\rm esc} = 100 $. {\it Lower right panel} (c): the case of different 
inter-shock distances, from the upper to the lower curve we have 
dashed-dotted line $L = 10$, solid line $L =20$ our fiducial case, long-dashed 
line $L = 50$, dotted line $L = 100$, short-dashed line $L= 10^4$.
The power-law spectrum corresponds to the single shock solution.}
\end{figure}

We now keep $\tilde{t}_{\rm esc}$ constant and equal to the fiducial value
of 100. For intershock distances $L \equiv \tilde{t}_{\rm adv} \gg 100$, 
no multiple shock effect is seen (Fig.~\ref{nmulti}c). As in the case of short
escape time (Fig.~\ref{nmulti}b), the stationary solution tends to that 
from a single shock. As $\tilde{t}_{\rm adv}$ decreases, the spectrum 
hardens and the spectral index can take all values between $3$ and $4$ 
(for $r=4$). This is consistent with our finding that $\tilde{t}_{\rm M}
\approx{\rm few}\times\tilde{t}_{\rm adv}$.
If the inter-shock distance is reduced to below one diffusion length 
$L<1$, the assumptions normally used to derive the basic transport 
equation (Skilling \cite{skilling75}) are violated. We have not investigated 
this regime; although it potentially interesting, since if the distribution remains 
almost isotropic, the diffusion approximation may in fact remain adequate.
For low momentum $p \le 1$, the spectrum steepens with decreasing $L$ and 
tends to $p^{0}$. 

\subsection{Multiple shock effect: the case with synchrotron losses}
The stationary spectrum computed with the fiducial parameters is given in 
Fig.~\ref{smulti}a. The inclusion of losses creates
a pile-up at a momentum where losses equal gains i.e., 
$\tilde{t}_{\rm M}=\tilde{t}_{\rm syn}$. From the simulation, we find 
this occurs at roughly $\tilde{p}=10^3$, which implies 
$\tilde{t}_{\rm M}\approx 100$ close to $5\times\tilde{t}_{\rm adv}$. This 
provides an independent check on the estimate made from the results presented in 
Figs.~\ref{nmulti}b and \ref{nmulti}c. The width of this hump on the low 
momentum side is determined by the momentum at which $\tilde{t}_
{\rm esc}=\tilde{t}_{\rm syn}$, where escape intervenes to overwhelm the 
effects of synchrotron losses. Since $\tilde{t}_{\rm esc}=50\tilde{t}_
{\rm adv}\approx10\tilde{t}_{\rm M}$, the hump makes its appearance about 
one order of magnitude before it peaks.

At momenta higher than the peak of the hump, the losses dominate over all 
other processes, since there $\tilde{t}_{\rm syn}<\tilde{t}_{\rm adv}
\le \tilde{t}_{\rm M}$, and, in this example, the acceleration at a single 
shock is approximately equal to the advection time (see Eq.~\ref{lagage}). 
Consequently, the spectrum cuts off exponentially. This conclusion is consistent 
with the results of Melrose \& Crouch~(\cite{melrosecrouch97}) found using an 
iterative method in which the synchrotron cut-off is estimated in advance. 

\subsubsection{Variations of the loss rate}
A decrease (increase) of $\tilde{a}_{\rm s}$ with $L$ and $\tilde{t}_{\rm esc}$ 
kept unchanged, leads to stronger (weaker) losses, and lower (higher) momenta 
at the peak of the pile-up. This is confirmed in Fig.~\ref{smulti}b, 
which also demonstrates that the peak momentum is directly proportional to the 
loss time $\tilde{t}_{\rm syn}$. The width of the hump is unaffected, 
since its lower bound also moves in proportion to $\tilde{t}_{\rm syn}$.
Radiative losses are unimportant below the injection momentum, except when 
the loss rate is so strong ($\tilde{a}_{\rm s} \sim 1$) that particle are 
prevented from being accelerated at all.

\subsubsection{Effects of the other parameters}
We stress here the effects of both escape and adiabatic losses on the 
shape of the pile-up. 

\begin{figure}[htbp]  
\centerline{\epsfxsize=10cm \epsfbox{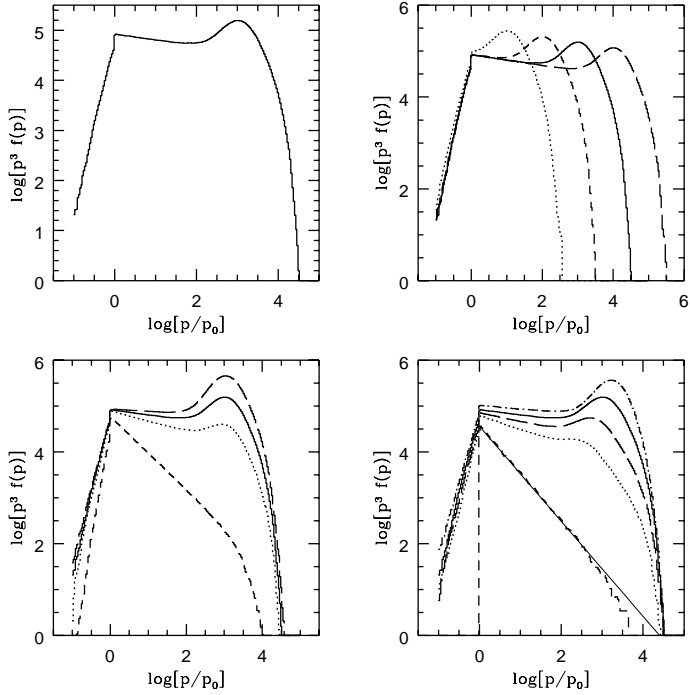}} 
\caption{
\label{smulti}
Differential logarithm spectrum for multiple shock 
acceleration with synchrotron losses. {\it Upper left panel} (a) our fiducial case, 
see text for the parameters. {\it Upper right panel} (b): the case with different 
synchrotron loss rates, the solid line corresponds to our refering case, 
the short dashed and dotted lines have a loss rate 
divided by 10 and 100, and the long dashed has a loss rate multiplied by 10.
{\it Lower left panel} (c): the case with different escape time, from the 
upper to lower curve: $\tilde{t}_{\rm esc} =$ $2 10^3$, $10^3$ our 
refering case, $500$, and $100$. {\it Lower right panel} (d): the case 
with different inter-shock distances, the distribution correponding to our 
refering value $L = 20$ is in solid line.
In dotted-dashed line $L=10$, in long-dashed line $L= 50$, in dotted line 
$L = 100$, in short-dashed line $L= 10^4$. The power-law spectrum 
corresponds to the single shock solution with $\tilde{t}_{\rm S}/\tilde{t}
_{\rm esc} \sim 2 10^{-2}$.}
\end{figure}

\begin{itemize}
\item Keeping the inter-shock distance $L \equiv \tilde{t}_{\rm adv}$ constant and 
allowing $\tilde{t}_{\rm esc}$ to vary causes a change in both the spectral slope
at low momenta, where losses are unimportant, and a change in the width
of the pile-up Fig.~\ref{smulti}c. As $\tilde{t}_{\rm esc}$ increases, the 
low momentum spectrum hardens, and the width of the hump increases, as it 
shortens, the spectrum softens, and the hump becomes less prominent, 
until it disappears completely once $\tilde{t}_{\rm esc}<\tilde{t}_{\rm adv}$. 
At this point, the power-law index of the low momentum spectrum is approximately 
$4$, in agreement with the findings of Kardashev~(\cite{kardashev62}).
   
\item Keeping the escape time constant, but varying the advection time
(and inter-shock distance) enables one to distinguish the regimes of 
single and multiple shock acceleration.
For larger values of $\tilde{t}_{\rm adv}$, than our fiducial case,
four regions in momentum space can be found. Starting at low momentum, 
these are (Fig.~\ref{smulti}d)
\begin{enumerate} 
\item 
For $\tilde{t}_{\rm syn}>\tilde{t}_{\rm esc}$, acceleration proceeds 
by the multiple shock process without losses.
\item
For $\tilde{t}_{\rm esc}>\tilde{t}_{\rm syn}>\tilde{t}_{\rm adv}$ the spectrum is 
affected by losses, and starts to form a pile-up. Acceleration at multiple 
shocks takes place.
\item
For $\tilde{t}_{\rm adv}>\tilde{t}_{\rm syn}>\tilde{t}_{\rm S}$ particles 
are prevented from reaching the next shock before cooling.
An interesting phenomenon appears in this case: the spectrum shows 
a power-law index appropriate to acceleration at a single shock front
(see the dotted line)
\item
For $\tilde{t}_{\rm S}>\tilde{t}_{\rm syn}$ losses dominate over all other 
processes and the spectrum cuts off.
\end{enumerate}
In Fig.~\ref{smulti}d all four regions can be distinguished in the case $L=100$.
At higher $L$, escape prevents multiple shock acceleration, and at lower 
values the shocks are too close together to allow the emergence of the 
single-shock power-law spectrum.
These effects do not appear in previous works, which had to assume a separation 
in the time-scales of the different. The high momentum power-law tail due to 
the single shock process may be important and contribute a substantial 
fraction of the total pattern luminosity.
\end{itemize}

\subsection{Synchrotron spectrum}
\begin{figure}[htbp]  
\centerline{\epsfxsize=8cm \epsfbox{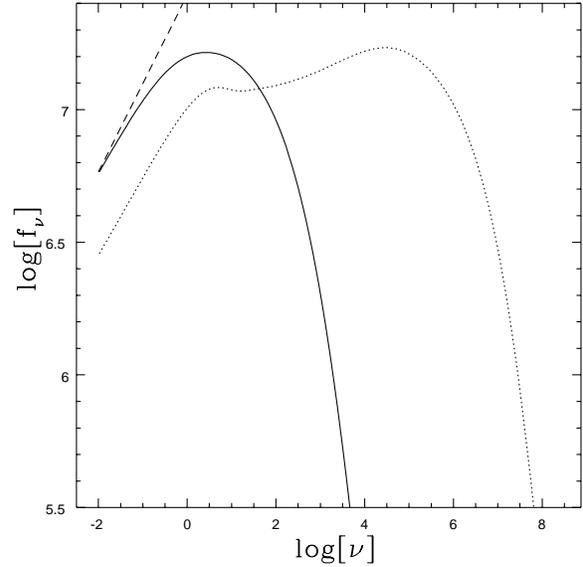}} 
\caption{
\label{synchro}
Synchrotron spectrum produced by a periodic pattern. 
The solid lines line shows the synchrotron radiations produced if the loss 
rate is one hundred times stronger (see the solution in dotted line of
figure 6 b). We compare also the low momentum slope with a $1/3$ power-law 
spectrum is dashed line. The dotted line curve is produced for an escape 
time a factor two larger than in our refering case (see the solution 
in long-dashed curve of figure 6 c).}
\end{figure}

The optically thin synchrotron emission produced by the 
particle distribution within a shock pattern
described by Fig.~\ref{pattern}, assuming constant magnetic 
field (i.e., parallel shocks) is given by 
\begin{equation}
S_{\nu} \propto \int_{\tilde{p}_{\rm min}}^{\tilde{p}_{\rm max}} \int_{-L/2}^{L/2} 
\tilde{P}(\nu, p) f(\tilde{p},\tilde{x}) \tilde{p}^2 \diff \tilde{p} 
\diff \tilde{x} \, .
\end{equation}
where, for a relativistic particle of charge $q$, mass $m$, and pitch-angle 
$\vartheta$,
\begin{equation}
\tilde{P}(\nu,p)= \left(\frac{\sqrt{3}}{2\pi}\right) 
\frac{q^3 B \sin\vartheta}{m c^2}
\frac{\nu}{\nu_c} \int_{\nu/\nu_{\rm c}}^{\infty} K_{5/3}(\xi)\diff\xi \, .
\end{equation}
The characteristic frequency depends on $p$ as $\nu_{\rm c}(p) = 3\omega_{\rm c}
\sin\vartheta \ p^2/(2m^2c^2) $, where the cyclotron frequency is
$\omega_{\rm c} = q B/ m c$. 
The integration over the particle position is done by computing the number
of particles at a given momentum $p$ at each time step inside the pattern. 
This gives the quantity $N(\tilde{p}) = \int_{-L/2}^{L/2} f(\tilde{p},\tilde{x}) 
\tilde{p}^2 \diff \tilde{p}$ 
contributing at a given frequency by the factor $\tilde{P}(\nu, p)$ 
to the synchrotron spectrum produced by the pattern. 
Figure~\ref{synchro} shows the resulting
synchrotron spectrum.

The results, shown in Fig.~\ref{synchro}, 
display a great variety of synchrotron spectra. For large 
ratios $t_{\rm esc}/t_{\rm adv}$ the spectrum is flat at low frequency, 
and extends over a range which depends on the relative 
strengths of loss and acceleration, which can be quite sufficient to 
explain the flat radio spectra observed in radio-loud quasars. 
Such an explanation provides an alternative to the inhomogeneous self-absorbed 
models usually advanced as the origin of flat radio Quasar spectra. 
Figure~\ref{synchro} also shows that inverted spectra 
over 2-3 decades are also quite possible, where the cooling time is still 
shorter than the escape time. In fact the pile-up effect, as stressed by 
Melrose (\cite{melrose96}), is essential to explain the inverted spectra 
observed in galactic centre sources such as Sagittarus $A^*$
(Beckert et al.~\cite{beckert96}), and also may be associated with the 
flaring states of some extragalactic sources, such as the recent 
X-ray bursts of Mrk~501 (Pian et al.~\cite{pianetal98}). 
Such flares can arise from a variation of either the inter-shock distance 
or the escape time from the system. A change in the magnetic field stength does
not mimick the multi-wavelength behaviour adequately 
(Mastichiadis \& Kirk~\cite{mastichiadiskirk97}).

\section{Conclusions and outlook}
\label{conclusions}
Achterberg \& Kr\"ulls (\cite{achterbergkruells92}) and
Kr\"ulls \& Achterberg~(\cite{KA94}, KA94)
have demonstrated the usefulness of the stochastic differential equation 
approach as an efficient tool for investigating 
particle acceleration at 
shock fronts. They found approximate solutions of the 
diffusion-advection equations in several different physical situations
including second order Fermi acceleration and momentum dependent spatial
diffusion coefficients. However, their computational approach is limited to 
spatially resolved shock structures, and needs 
unrealistically short time steps for
thin shocks. We have presented an improved scheme
in which the particle momentum gain during a time step 
depends on both the initial and final positions of the particle.
The scheme reproduces analytical results derived for shock thicknesses much 
lower than a typical diffusive length, without imposing an excessive 
requirement on the time step.
We have applied our procedure to a system of multiple identical 
shocks. The $p^{-3}$ signature is obtained when the escape time is much
larger than the multiple shock acceleration time. Inclusion of losses
leads to a pile-up effect where the acceleration and loss rates are equal,
as has been found analytically.
The pile-up is present only for a restricted range of parameters: 
when the escape time exceeds the time for the plasma to move from one shock to 
the next, but is short enough not to permit synchrotron cooling of the lowest 
energy particles. We have shown that parameter ranges can be found in which 
the spectrum simultaneously displays the power law index characteristic of 
multiple shock acceleration (at low momenta) and the index appropriate 
to single shock acceleration at high momenta.
  
The synchrotron spectrum produced in each flow pattern can show flat or 
inverted spectra depending on the inter-shock distance, the loss strength and 
the ratio of escape to acceleration time. Multiple 
shock acceleration may explain the radio spectrum from flat radio Quasars, 
the radio to IR spectrum of galactic sources such as Sagittarus $A^*$, 
and even hard X-ray spectra observed in high states of the BL Lac 
Markarian 501.

Several extensions of this work are possible, such as the consideration of 
non-periodic flows and non-stationary injection. It also seems feasible to
extend the 1D simulations to more complex multi-dimensional flows, such as
numerical simulations of jets.

\begin{acknowledgements}
This work was supported by the European Commission under the 
TMR Programme, contract number FMRX-CT98-0168. The authors thank Prof A. Achterberg
for fruitful discussions.
\end{acknowledgements}

\end{document}